Letter

# The neural networks with tensor weights and emergent fermionic Wick rules in the large-width limit

Guojun Huang [a,b], Kai Zhou [a,c,*]

[a] The School of Science and Engineering, The Chinese University of Hong Kong (Shenzhen), Longgang, Shenzhen, Guangdong, 518172, China
[b] Physics Department, Tsinghua University, Beijing, 100084, China
[c] The School of Artificial Intelligence, The Chinese University of Hong Kong (Shenzhen), Longgang, Shenzhen, Guangdong, 518172, China

ABSTRACT

In this paper, we study complex-valued neural network (CVNNs) with tensor-valued hidden-to-output weights within the framework of neural-network quantum field theory (NN-QFT). For standard CVNNs with scalar weights, we derive the generating functional and identify the exact Gaussian process that arises in the infinite-width limit, together with its associated effective quantum state. When the last-layer weights are promoted to Clifford-algebra-valued tensors, the network output becomes complex matrix-valued, and a fermion-like sign structure in the large-width correlation functions of the network output is induced. We show that, in the infinite-width limit, correlators with equal numbers of $f^\dagger$ and $f$ obey fermionic Wick rules and can be written as determinants built from a scalar Euclidean kernel $S(x,y) = \langle f^\dagger(x) f(y)\rangle$. This provides a sign-structured extension of NN-QFT at the level of Euclidean correlators and Feynman rules, even though a microscopic Grassmann path integral representation for the network parameters has not yet been constructed. Our analysis thus pushes the NN-QFT correspondence beyond purely bosonic Gaussian fields and suggests a possible route to encoding fermion-like symmetries in neural architectures for QFT correspondence.

## 1. Introduction

Wide neural networks with i.i.d. random Gaussian weights have been rigorously connected to free bosonic quantum field theories, forming a one-to-one correspondence now known as neural-network quantum field theory (NN-QFT) [1–14]. For standard feed-forward architectures with real-valued parameters, it has been shown [1,4] that the $n$-th order correlation function of the network output can be expressed as a sum over bosonic $n$-points Feynman diagrams,

$$G^{(n)}_{\rm NN}(\boldsymbol{x}_1, \boldsymbol{x}_2, \dots, \boldsymbol{x}_n) \equiv \frac{1}{n_{\rm nets}} \sum_{\alpha=1}^{n_{\rm nets}} f_\alpha(\boldsymbol{x}_1) f_\alpha(\boldsymbol{x}_2) \cdots f_\alpha(\boldsymbol{x}_n), \tag{1}$$

where $f_\alpha(\boldsymbol{x}_n)$ denotes the network output evaluated at the space-time position $\boldsymbol{x}_n$ for a particular realisation of the randomly sampled weights $\alpha$. For the single-hidden-layer architecture, it can be written as

$$f_\alpha(\boldsymbol{x}_j) = \sigma(\boldsymbol{x}_j W^{(\alpha)}_{\rm in,h} + b^{(\alpha)}_{\rm h}) W^{(\alpha)}_{\rm h,out}, \tag{2}$$

where $W_{\rm in,h}$ and $W_{\rm h,out}$ are, respectively, the input-to-hidden and hidden-to-output weight matrices, $b_h$ is the bias of the hidden layer, the subscripts $h$ label hidden neurons while superscripts $\alpha$ label independent network realisations, and $\sigma$ is the element-wise activation function. In this work we adopt the exponential activation [1]

$$\sigma(\boldsymbol{x}_j W_{\rm in,h} + b_{\rm h}) = \frac{\exp(\boldsymbol{x}_j W_{\rm in,h} + b_{\rm h})}{\exp\left(\sigma_{\rm b}^2 + \sigma_{\rm w}^2 \boldsymbol{x}_j^2 / D\right)}, \tag{3}$$

which normalizes the mean pre-activation to zero under the Gaussian weight and bias ensemble.

Pushing the hidden-layer width $H$ to infinity collapses the diagrammatic expansion to its Gaussian core: all interaction vertexes are suppressed by factors of $\mathcal{O}(1/H)$, so the network's effective Lagrangian reduces to that of a free scalar field and only the two-point propagator survives [12]. In this asymptotic regime the model is mathematically equivalent to a Gaussian process, yet the precise quantum state that emerges from the weight ensemble–the field configuration measure that reproduces these correlators–has not been fully characterized.

When the hidden layer is finite yet parametrically wide, $H \gg 1$, the leading $1/H$ corrections to the free Lagrangian organize themselves into a low-energy effective theory [4]. A systematic Taylor expansion of the activation kernel reveals that the first non-vanishing interaction is quartic, so the network's dynamics are governed by a $\phi^4$ term whose coupling scales as $\mathcal{O}(H^{-1})$. Because the hidden-to-output weights $W_{\rm h,out}$ are drawn from an even distribution, all odd powers cancel, leaving this quartic vertex as the dominant interaction mode.

* Corresponding author.
*E-mail address:* zhoukai@cuhk.edu.cn (K. Zhou).

For neural networks with genuinely narrow hidden layers, the $\phi^4$ approximation fails: higher-order vertexes proliferate and the diagrammatic expansion of the $n$-point correlators becomes markedly intricate. In this regime, the weight ensemble generates an entire hierarchy of interaction terms beyond quartic order, rendering perturbative treatments intractable.

The bosonic nature of the field $f$ is determined by the distribution of $W_{\mathrm{h,out}}$, to which the i.i.d. real-Gaussian distribution can induce the non-interaction Bose field. A central obstraction is that fermionic statistics cannot be obtained by a naive reshaping of the bosonic weight ensemble, since the Lagrangian of fermionic field can not simply be connected to the distribution. Replacing each scalar weight component by tensor-valued coefficients [15] that form a basis of Clifford generators, $\{\gamma_h\}$, forces every weight component to anticommute with every other. This single architectural change converts the Gaussian process that arises at $H \to \infty$ from Bose to Fermi statistics while leaving the activation kernel, the diagrammatic rules and the large-$H$ power counting untouched. Complex-valued networks provide the minimal setting for such tensor weights, and are thus prospective to be connected to fermionic field theory.

While previous works focused on bosonic fields, here we establish the first explicit neural-construction in NN-QFT correspondence shows fermionic Wick structured correlators.

Earlier work on machine learning fermionic systems has focused on using neural networks to represent many-body wavefunctions of a prescribed Hamiltonian [16–21]. Here we tackle the complementary questions posed by NN-QFT: **what neural architecture gives rise *intrinsically* to a fermionic quantum field theory?** We show that complex-valued networks equipped with Clifford-tensor output weights answer this question positively–providing, in effect, a learnable sign-structured surrogate whose large-width correlators follow fermionic Wick combinatorics. This closes a conceptual gap in the NN-QFT programme at the level of correlators, extending it beyond the purely bosonic domain explored to date and providing a sign-structured surrogate for free complex fermion correlators. This work also joins a growing synergy between physics and machine learning, where neural networks have been applied to challenges in physics (e.g., condensed matter physics [22–26] and high energy nuclear physics [27–31]). Conversely, physical principles have inspired novel machine learning developments [32,33] as well, demonstrating rich cross-disciplinary enrichment.

The rest of the paper is organized as follows. Section 2 derives the generating functional of complex-valued neural networks and interprets it as a path integral for a complex scalar field. Section 3 obtains the exact quantum state for the infinite hidden layer width limit, identifying the input-to-hidden parameters as eigenvalues and the hidden-to-output parameters as dynamical fields.

Section 4 promotes those dynamical fields to Clifford tensors, thereby realizing a fermionic theory and demonstrating how anticommutation arises diagrammatically. Finally, Section 5 summarizes the results and outlines future directions.

## 2. Generating functional of CVNN

To extend the NN-QFT correspondence into the complex domain, we upgrade every parameter of the network to the complex plane. In the resulting complex-valued neural network (CVNN) [34–41], each weight and bias decomposes into statistically independent real and imaginary parts, both sampled from zero-mean Gaussian ensemble [1]:

$$\mathrm{Re}(W^{(\alpha)}_{\mathrm{in,h}}),\ \mathrm{Im}(W^{(\alpha)}_{\mathrm{in,h}}) \sim \mathcal{N}(\mu_{\mathrm{w}}, \sigma_{\mathrm{w}}/\sqrt{D}), \tag{4}$$

$$\mathrm{Re}(W^{(\alpha)}_{\mathrm{h,out}}),\ \mathrm{Im}(W^{(\alpha)}_{\mathrm{h,out}}) \sim \mathcal{N}(\mu_{\mathrm{w}}, \sigma_{\mathrm{w}}/\sqrt{H}), \tag{5}$$

$$\mathrm{Re}(b^{(\alpha)}_{\mathrm{h}}),\ \mathrm{Im}(b^{(\alpha)}_{\mathrm{h}}) \sim \mathcal{N}(\mu_{\mathrm{b}}, \sigma_{\mathrm{b}}). \tag{6}$$

Ensemble averages over these Gaussians can be interpreted as a discrete path integral in the hidden-unit index $h$: the product over $h = 1, \dots, H$ forms a lattice of "time slices", and in the continuum limit $H \to \infty$ it recovers the usual time-ordered functional integral that underlies quantum field theory.

### 2.1. The path integral explanation

To access the $n$-points correlation function, we couple the network to an auxiliary source field $J(\boldsymbol{x})$ and work with a generating functional. In the condition that the $f_\alpha$ is a scalar field, one can denote $W_{\mathrm{in,h}}$, $b_{\mathrm{h}}$ and $W_{\mathrm{h,out}}$ with $\boldsymbol{Q}_h$, $V_h$ and $\varphi_h$ as

$$\boldsymbol{Q}_h \equiv W_{\mathrm{in,h}}, \tag{7}$$

$$V_h \equiv b_{\mathrm{h}}, \tag{8}$$

$$\varphi_h \equiv H * W_{\mathrm{h,out}}, \tag{9}$$

with the dimension of vector $\boldsymbol{Q}_h$ being $D$. Then the output function of the CVNN becomes

$$f(\boldsymbol{Q}, V|\boldsymbol{x}) = \frac{1}{H}\sum_{h=1}^{H} \lambda_h(\boldsymbol{Q}, V|\boldsymbol{x})\varphi_h, \tag{10}$$

$$\lambda_h(\boldsymbol{Q}, V|\boldsymbol{x}) \equiv \frac{\exp\left(\boldsymbol{x}\cdot\boldsymbol{Q}_h + V_h\right)}{\exp\left(\sigma_{\mathrm{b}}^2 + \sigma_{\mathrm{w}}^2\boldsymbol{x}^2/D\right)}, \tag{11}$$

where $\lambda_h$ plays the role of a hopping amplitude between the hidden-unit "sites" indexed by $h$ and continuum coordinate $x$. The generating functional is therefore written as a path integral [42]:

$$\begin{aligned} &Z[J^*, J] \\ &= \prod_{h=1}^{H}\left\{\int \mathrm{d}\boldsymbol{Q}_h \mathrm{d}\boldsymbol{Q}_h^* \mathrm{d}V_h \mathrm{d}V_h^* \mathrm{d}\varphi_h \mathrm{d}\varphi_h^* \right.\\ &\left.\times \frac{\mathrm{e}^{-||\boldsymbol{Q}_h||^2/(2\sigma_{\mathrm{w}}^2/D)}}{(2\pi\sigma_{\mathrm{w}}^2/D)^D}\frac{\mathrm{e}^{-|V_h|^2/(2\sigma_{\mathrm{b}}^2)}}{2\pi\sigma_{\mathrm{b}}^2}\frac{\mathrm{e}^{-|\varphi_h|^2/(2\sigma_{\mathrm{w}}^2 H)}}{2\pi\sigma_{\mathrm{w}}^2 H}\right\} \\ &\times \mathrm{e}^{\mathrm{i}\int \mathrm{d}^D\boldsymbol{x}\ [J(\boldsymbol{x})f^*(\boldsymbol{Q},V|\boldsymbol{x}) + J^*(\boldsymbol{x})f(\boldsymbol{Q},V|\boldsymbol{x})]}, \end{aligned} \tag{12}$$

the correlation functions of $f$ can be obtained via calculating the functional derivatives of the auxiliary field $J(\boldsymbol{x})$ and $J^*(\boldsymbol{x})$. The $\varphi_h$ integration is easy to handle for being the linear combination, according to the integral formula that

$$\begin{aligned} \mathrm{e}^{-2\sigma_{\mathrm{w}}^2|C|^2/H} = &\int \frac{\mathrm{d}\varphi_{\mathrm{R}}\mathrm{d}\varphi_{\mathrm{I}}}{2\pi\sigma_{\mathrm{w}}^2 H}\mathrm{e}^{-(\varphi_{\mathrm{R}}^2+\varphi_{\mathrm{I}}^2)/(2\sigma_{\mathrm{w}}^2 H)} \\ &\times \mathrm{e}^{\mathrm{i}C(\varphi_{\mathrm{R}}-\mathrm{i}\varphi_{\mathrm{I}})/H + \mathrm{i}C^*(\varphi_{\mathrm{R}}+\mathrm{i}\varphi_{\mathrm{I}})/H}. \end{aligned} \tag{13}$$

The generating functional then can be written as the expectation over the $\boldsymbol{Q}_h$, $V_h$ distribution as

$$Z[J^*, J] = \mathbb{E}_{\boldsymbol{Q},V}\left[\exp\left(\frac{2\sigma_{\mathrm{w}}^2}{H}\sum_{h=1}^{H}\tilde{S}_h(\boldsymbol{Q}, V)\right)\right], \tag{14}$$

where we denote the effective action $\tilde{S}_h(\boldsymbol{Q}, V)$ as

$$\tilde{S}_h(\boldsymbol{Q}, V) \equiv -\int \mathrm{d}^D\boldsymbol{x}\mathrm{d}^D\boldsymbol{y} J(\boldsymbol{x})\mathcal{M}_h(\boldsymbol{Q}, V|\boldsymbol{x}, \boldsymbol{y})J^*(\boldsymbol{y}), \tag{15}$$

with the kernel being defined as $\mathcal{M}_h(\boldsymbol{Q}, V|\boldsymbol{x}, \boldsymbol{y}) \equiv \lambda_h^*(\boldsymbol{Q}, V|\boldsymbol{x})\lambda_h(\boldsymbol{Q}, V|\boldsymbol{y})$. When considering the wide width hidden layer condition, one can divide the generating functional by different $h$, the powers of $\tilde{S}_h(\boldsymbol{Q}, V)$ term can be expanded by orders of $1/H$ as

$$Z[J^*, J] = \prod_{h=1}^{H}\left\{\sum_{m=0}^{\infty}\frac{1}{m!}\left(\frac{2\sigma_{\mathrm{w}}^2}{H}\right)^m \mathbb{E}_{\boldsymbol{Q},V}[(\tilde{S}_h(\boldsymbol{Q}, V))^m]\right\}. \tag{16}$$

Considering that the expectation $\mathbb{E}_{\boldsymbol{Q},V}[(\tilde{S}_h(\boldsymbol{Q}, V))^m]$ gives the same value for arbitrary $h$, one can replace it with the value at $h = 1$ as $\mathbb{E}_{\boldsymbol{Q},V}[(\tilde{S}_1(\boldsymbol{Q}, V))^m]$. Then, the generating functional can be simplified as

$$Z[J^*, J] = \prod_{h=1}^{H} Z_h[J^*, J] = (Z_1[J^*, J])^H, \tag{17}$$

with $Z_1[J^*, J]$ being

$$
\begin{aligned}
&Z_1[J^*, J] \\
&=1+\sum_{m=1}^{\infty}\frac{(-1)^m}{m!}\frac{(2\sigma_{\mathrm{w}}^2)^m}{H^m}\int\prod_{i=1}^{m}[\mathrm{d}^D\boldsymbol{x}_i\mathrm{d}^D\boldsymbol{y}_i]\left[\prod_{j=1}^{m}J(\boldsymbol{x}_j)\right] \\
&\times\mathbb{E}_{\boldsymbol{Q},V}\left[\prod_{k=1}^{m}\mathcal{M}_1(\boldsymbol{Q},V|\boldsymbol{x}_k,\boldsymbol{y}_k)\right]\left[\prod_{l=1}^{m}J^*(\boldsymbol{y}_l)\right].
\end{aligned} \tag{18}
$$

The coefficients of the expansion actually represents the Feynman diagrams, and it can be noticed that the formation of the above generating functional is not the traditional version, $Z[J^*, J] \sim \mathrm{e}^{\mathrm{i}W[J^*,J]}$, with $W[J^*, J]$ being the connect generating functional. One can still use the formula $W[J^*, J] = -\mathrm{i}\ln Z[J^*, J]$ to calculate the connect generating functional, however, the result is not simple for narrow width hidden layer condition, but in the wide width hidden layer condition, one can keep the lowest contribution beyond Gaussian process, and obtain the effective connect generating functional.

### 2.2. The connected generating functional

In the wide-width regime ($H \gg 1$) the product over hidden units factorises, and the familiar limitation formula

$$\lim_{H\to\infty}(1+\lambda/H)^H = \mathrm{e}^{\lambda}, \tag{19}$$

allows us to resum the leading contributions. Expanding the single-unit effective action ($\tilde{S}_1(Q, V)$) in powers of the source and then applying the limit gives the cumulant series

$$\lambda=\sum_{m=1}^{\infty}\frac{1}{m!}\frac{(2\sigma_{\mathrm{w}}^2)^m}{H^{m-1}}\mathbb{E}_{\boldsymbol{Q},V}[(\tilde{S}_1(\boldsymbol{Q},V))^m], \tag{20}$$

so that, to quartic order in the source, the connected generating functional can then be approximated as

$$
\begin{aligned}
&\ln Z[J^*, J] \\
&\approx -2\sigma_{\mathrm{w}}^2\int\mathrm{d}^D\boldsymbol{x}\int\mathrm{d}^D\boldsymbol{y}\;J(\boldsymbol{x})\mathrm{e}^{-\frac{\sigma_{\mathrm{w}}^2}{D}(\boldsymbol{x}-\boldsymbol{y})^2}J^*(\boldsymbol{y}) \\
&+\frac{2\sigma_{\mathrm{w}}^4}{H}\int\mathrm{d}^D\boldsymbol{x}_1\int\mathrm{d}^D\boldsymbol{x}_2\int\mathrm{d}^D\boldsymbol{y}_1\int\mathrm{d}^D\boldsymbol{y}_2 \\
&\times J(\boldsymbol{x}_1)J(\boldsymbol{x}_2)\Lambda(\boldsymbol{x}_1,\boldsymbol{x}_2,\boldsymbol{y}_1,\boldsymbol{y}_2)J^*(\boldsymbol{y}_1)J^*(\boldsymbol{y}_2)
\end{aligned} \tag{21}
$$

with the quartic kernel $\Lambda(\boldsymbol{x}_1, \boldsymbol{x}_2, \boldsymbol{y}_1, \boldsymbol{y}_2)$ being

$$
\begin{aligned}
&\Lambda(\boldsymbol{x}_1,\boldsymbol{x}_2,\boldsymbol{y}_1,\boldsymbol{y}_2) \\
&\equiv\mathbb{E}_{\boldsymbol{Q},V}\left[\mathcal{M}_1(\boldsymbol{Q},V|\boldsymbol{x}_1,\boldsymbol{y}_1)\mathcal{M}_1(\boldsymbol{Q},V|\boldsymbol{x}_2,\boldsymbol{y}_2)\right].
\end{aligned} \tag{22}
$$

This approximation shows that, once the hidden layer is merely parametrically wide, the leading connected diagrams organise into an effective $\phi^4$ theory whose coupling is suppressed by $1/H$. This formulation is more transparent for CVNN-QFT than attempting to approximate the Lagrangian directly at quartic order, because it keeps track of the finite-width correctons that distinguish the network from its strict $H \to \infty$ Gaussian-process limit.

### 2.3. The renormalization of finite width hidden layer CVNN

To analyse ultraviolet (UV) effects we switch to momentum space and impose a hard cutoff [42,43].

Consider the Fourier transformation of low momentum mode $f_<(\boldsymbol{Q}, V|\boldsymbol{x})$ of the network output,

$$f_<(\boldsymbol{Q},V|\boldsymbol{x})\equiv\int\frac{\mathrm{d}^D\boldsymbol{p}}{(2\pi)^{D/2}}\mathrm{e}^{-\mathrm{i}\boldsymbol{x}\cdot\boldsymbol{p}}f_<(\boldsymbol{Q},V|\boldsymbol{p}), \tag{23}$$

with $f_<(\boldsymbol{Q}, V|\boldsymbol{p})$ being restricted to integral area $\mathcal{V}$ as

$$f_<(\boldsymbol{Q},V|\boldsymbol{p})\equiv\begin{cases}f(\boldsymbol{Q},V|\boldsymbol{p}), & \boldsymbol{p}\in\mathcal{V},\\ 0, & \text{others},\end{cases} \tag{24}$$

where $\mathcal{V}$ is the area of $D$-dimensional cube of side length $p_L$ (thus $p_i \in [-p_L/2, p_L/2]$, $(i = 1, 2, \ldots, D)$) and $f(\boldsymbol{Q}, V|\boldsymbol{p})$ is the momentum representation of $f(\boldsymbol{Q}, V|\boldsymbol{x})$,

$$f(\boldsymbol{Q},V|\boldsymbol{p})=\frac{1}{H}\sum_{h=1}^{H}\lambda_h(\boldsymbol{Q},V|\boldsymbol{p})\varphi_h, \tag{25}$$

$$\lambda_h(\boldsymbol{Q},V|\boldsymbol{p})\equiv\left(\frac{D}{2\sigma_{\mathrm{w}}^2}\right)^{D/2}\mathrm{e}^{-\frac{(\boldsymbol{p}-\mathrm{i}\boldsymbol{Q}_h)^2}{4\sigma_{\mathrm{w}}^2/D}+V_h-\sigma_{\mathrm{b}}^2}. \tag{26}$$

It can be noticed that the UV divergence arises when we calculate the Feynman diagrams in momentum representation. This divergence is caused by exchanging the integral order of $\boldsymbol{Q}$ (especially the real part of $\boldsymbol{Q}$) and $\boldsymbol{p}$, mathematically. Keeping the original integration order removes the divergence and the coefficients in generating functional Eq. (18) transforms to the ones in integral area $\mathcal{V}$ via replacing $\mathcal{M}_1(\boldsymbol{Q}, V|\boldsymbol{x}_k, \boldsymbol{y}_k)$ with its low-momentum counterpart $\mathcal{M}_1^<(\boldsymbol{Q}, V|\boldsymbol{x}_k, \boldsymbol{y}_k)$, where $\mathcal{M}_h^<(\boldsymbol{Q}, V|\boldsymbol{x}, \boldsymbol{y})$ is defined as

$$\mathcal{M}_h^<(\boldsymbol{Q},V|\boldsymbol{x},\boldsymbol{y})\equiv\lambda_h^{<*}(\boldsymbol{Q},V|\boldsymbol{x})\lambda_h^<(\boldsymbol{Q},V|\boldsymbol{y}), \tag{27}$$

$$\lambda_h^<(\boldsymbol{Q},V|\boldsymbol{x})\equiv\int_{\boldsymbol{p}\in\mathcal{V}}\frac{\mathrm{d}^D\boldsymbol{p}}{(2\pi)^{D/2}}\mathrm{e}^{-\mathrm{i}\boldsymbol{p}\cdot\boldsymbol{x}}\lambda_h(\boldsymbol{Q},V|\boldsymbol{p}). \tag{28}$$

A single Wilsonian RG step enlarges the cutoff, $p_L \mapsto s\,p_L$) with $s > 1$. Under this rescaling the connected $m$-point coefficients transform as

$$\mathbb{E}_{\boldsymbol{Q},V}\left[\prod_{k=1}^{m}\mathcal{M}_1^<(\boldsymbol{Q},V|\boldsymbol{x}_k,\boldsymbol{y}_k)\right]\Rightarrow\mathbb{E}_{s\boldsymbol{Q},V}\left[\prod_{k=1}^{m}\mathcal{M}_1^<(\boldsymbol{Q},V|s\boldsymbol{x}_k,s\boldsymbol{y}_k)\right], \tag{29}$$

where the notation $\mathbb{E}_{s\boldsymbol{Q},V}$ means that the variance of each real component of $\boldsymbol{Q}_h$ is rescaled, $\sigma_w^2/D \mapsto \sigma_w^2/(s^2 D)$.

Taking $s \to \infty$ restores the original uncut theory:

$$
\begin{aligned}
&\lim_{s\to+\infty}\mathbb{E}_{s\boldsymbol{Q},V}\left[\prod_{k=1}^{m}\mathcal{M}_1^<(\boldsymbol{Q},V|s\boldsymbol{x}_k,s\boldsymbol{y}_k)\right] \\
&=\mathbb{E}_{\boldsymbol{Q},V}\left[\prod_{k=1}^{m}\mathcal{M}_1(\boldsymbol{Q},V|\boldsymbol{x}_k,\boldsymbol{y}_k)\right],
\end{aligned} \tag{30}
$$

hence the coefficients of the full CVNN-QFT are invariant under the scaling $\boldsymbol{x} \to s\,\boldsymbol{x}$ once the cutoff has been removed, confirming that they define a fixed-point theory

$$\mathbb{E}_{\boldsymbol{Q},V}\left[\prod_{k=1}^{m}\mathcal{M}_1(\boldsymbol{Q},V|\boldsymbol{x}_k,\boldsymbol{y}_k)\right]=\mathbb{E}_{s\boldsymbol{Q},V}\left[\prod_{k=1}^{m}\mathcal{M}_1(\boldsymbol{Q},V|s\boldsymbol{x}_k,s\boldsymbol{y}_k)\right]. \tag{31}$$

The hard-cutoff procedure therefore provides a consistent renormalisation scheme for a finite-width CVNN. It eliminates spurious UV divergences at each intermediate scale while preserving the large-width expansion that underpins the $\phi^4$ effective description derived in above.

Note that our large-width expansion assumes i.i.d. last-layer weights with finite moments required to define the kernel and suppress higher vertices. The choice of exponential activation is made for analytic tractability of the CVNN generating functional; the fermion-like Wick combinatorics in Section IV relies primarily on the last-layer Clifford structure.

## 3. The quantum state in the infinite-width limit

In this section, we study the infinite-width hidden layer NN, $H \to \infty$, and extract the corresponding quantum state of the network in this limit. The $h$ summation can be regarded as the $H$-times sampling of the Gaussian distributions $\boldsymbol{Q}_h \sim \mathcal{N}(0, \sigma_{\mathrm{w}}/\sqrt{D})$ and $V_h \sim \mathcal{N}(0, \sigma_{\mathrm{b}})$, then in the infinite-width hidden layer limit, the summation becomes the expectation of $\boldsymbol{Q}$ and $V$ with respect to their distributions, which via the Riemann summation method can be written as a Riemann integral over a dummy variable $\xi \in [0, 1]$:

$$f_\infty(\boldsymbol{Q},V|\boldsymbol{x})=\int_0^1\mathrm{d}\xi\frac{\mathrm{e}^{\boldsymbol{x}\cdot\boldsymbol{Q}(\xi)+V(\xi)}}{\mathrm{e}^{\sigma_{\mathrm{b}}^2+\sigma_{\mathrm{w}}^2\boldsymbol{x}^2/D}}\varphi(\xi), \tag{32}$$

the action of $f$ becomes the free field action, because the vertexes higher than 4-points (including) vanish for measuring zero. For example, denote operator $\hat{O}_h(\boldsymbol{x}) \equiv |\lambda_h(\boldsymbol{Q}, V|\boldsymbol{x})|^2 = \lambda_h^*(\boldsymbol{Q}, V|\boldsymbol{x})\lambda_h(\boldsymbol{Q}, V|\boldsymbol{x})$, the 4-points correlation function gives

$$\left\langle |f(\boldsymbol{Q}, V|\boldsymbol{x})|^4 \right\rangle = 8\sigma_{\mathrm{w}}^4 \left\{\mathbb{E}_{\boldsymbol{Q},V}\left[\hat{O}_1(\boldsymbol{x})\right]\right\}^2 + \frac{8\sigma_{\mathrm{w}}^4}{H}\mathbb{E}_{\boldsymbol{Q},V}\left[(\Delta\hat{O}_1(\boldsymbol{x}))^2\right], \tag{33}$$

with the fluctuation operator defined as $\Delta\hat{O}_h(\boldsymbol{x}) = \hat{O}_h(\boldsymbol{x}) - \mathbb{E}_{\boldsymbol{Q},V}[\hat{O}_h(\boldsymbol{x})]$, this shows that the quantum fluctuations are depressed by the measuring zero under the limitation $H \to \infty$. This shows that $f_\infty(\boldsymbol{Q}, V|\boldsymbol{x})$ has the Gaussion like distribution, with its variance being the familiar neural-network kernel

$$\mathbb{E}_{\boldsymbol{Q},V}\left[\lambda_\xi^*(\boldsymbol{Q}, V|\boldsymbol{x})\lambda_\xi(\boldsymbol{Q}, V|\boldsymbol{y})\right] = \mathrm{e}^{-\frac{\sigma_{\mathrm{w}}^2}{D}(\boldsymbol{x}-\boldsymbol{y})^2}. \tag{34}$$

Since the higher order correlation functions vanish in the $H \to \infty$ limit, we can absorb the arithmetic square root of $\boldsymbol{Q}(\xi)$ and $V(\xi)$'s distributions into the definition of the field and write an equivalent path-integral representation, so that the 2-pt correlation function of $f$ equivalently reproduces the expectation of $\left\langle f^*(\boldsymbol{Q}, V|\boldsymbol{x})f(\boldsymbol{Q}, V|\boldsymbol{y})\right\rangle$,

$$f_{\mathrm{eff}}(\boldsymbol{Q}, V|\boldsymbol{x}) = \int_0^1 \mathrm{d}\xi \frac{\mathrm{e}^{-|V(\xi)|^2/(4\sigma_{\mathrm{b}}^2)}}{\sqrt{2\pi\sigma_{\mathrm{b}}^2}} \frac{\mathrm{e}^{-||\boldsymbol{Q}(\xi)||^2/(4\sigma_{\mathrm{w}}^2/D)}}{(2\pi\sigma_{\mathrm{w}}^2/D)^{D/2}} \times \frac{\exp\left(\boldsymbol{x}\cdot\boldsymbol{Q}(\xi) + V(\xi)\right)}{\exp\left(\sigma_{\mathrm{b}}^2 + \sigma_{\mathrm{w}}^2\boldsymbol{x}^2/D\right)}\varphi(\xi), \tag{35}$$

it can be simplified as

$$\begin{aligned} &f_{\mathrm{eff}}(\boldsymbol{Q}, V|\boldsymbol{x}) \\ &= \int_0^1 \mathrm{d}\xi \frac{\mathrm{e}^{-|V(\xi)-2\sigma_{\mathrm{b}}^2|^2/(4\sigma_{\mathrm{b}}^2)-\frac{1}{2}[V^*(\xi)-V(\xi)]}}{\sqrt{2\pi\sigma_{\mathrm{b}}^2}} \\ &\times \frac{\mathrm{e}^{-||\boldsymbol{Q}(\xi)-2\sigma_{\mathrm{w}}^2\boldsymbol{x}/D||^2/(4\sigma_{\mathrm{w}}^2/D)-\frac{1}{2}\boldsymbol{x}\cdot[\boldsymbol{Q}^*(\xi)-\boldsymbol{Q}(\xi)]}}{(2\pi\sigma_{\mathrm{w}}^2/D)^{D/2}}\varphi(\xi), \end{aligned} \tag{36}$$

the $V^*(\xi) - V(\xi)$ term has no effect for they vanishes in the Wick contraction of mutually conjugate fields $\left\langle f_{\mathrm{eff}}^*(\boldsymbol{Q}, V|\boldsymbol{x})f_{\mathrm{eff}}(\boldsymbol{Q}, V|\boldsymbol{y})\right\rangle$, besides, the integral of $V(\xi)$ makes no difference considering its integral transformation $V(\xi) \to \tilde{V}(\xi) \equiv V(\xi) + 2\sigma_{\mathrm{b}}^2$ for arbitrary $\boldsymbol{x}$, namely, the parameter $V(\xi)$ can be ignored and $f_{\mathrm{eff}}(\boldsymbol{Q}, V|\boldsymbol{x})$ can be equivalently written as

$$\tilde{f}_{\mathrm{eff}}(\boldsymbol{Q}|\boldsymbol{x}) = \int_0^1 \mathrm{d}\xi \frac{\mathrm{e}^{-\frac{[\boldsymbol{Q}_{\mathrm{R}}(\xi)-2\sigma_{\mathrm{w}}^2\boldsymbol{x}/D]^2+\boldsymbol{Q}_{\mathrm{I}}(\xi)^2}{4\sigma_{\mathrm{w}}^2/D}+\mathrm{i}\boldsymbol{x}\cdot\boldsymbol{Q}_{\mathrm{I}}(\xi)}}{(2\pi\sigma_{\mathrm{w}}^2/D)^{D/2}}\varphi(\xi). \tag{37}$$

The Wick contraction of $\varphi(\xi)$ actually controls the equivalence of $\boldsymbol{Q}(\xi)$, and one can transform $\varphi$ to the function of $\boldsymbol{Q}$ with its 2-pt correlation function gives the Dirac delta function of $\boldsymbol{Q}$, namely,

$$\varphi(\xi) \to \tilde{\varphi}(\boldsymbol{Q}), \tag{38}$$

with $\langle\tilde{\varphi}^*(\boldsymbol{Q})\tilde{\varphi}(\boldsymbol{Q}')\rangle = 2\sigma_{\mathrm{w}}^2\delta^{(D)}(\boldsymbol{Q} - \boldsymbol{Q}')$. Then the field $f$ in the $H \to \infty$ can be represented by the summation of the Gaussian wave as

$$\tilde{f}_{\mathrm{eff}}(\boldsymbol{x}) = \int \frac{\mathrm{d}^D\boldsymbol{Q}_{\mathrm{R}}\mathrm{d}^D\boldsymbol{Q}_{\mathrm{I}}}{(2\pi\sigma_{\mathrm{w}}^2/D)^{D/2}} \mathrm{e}^{-\frac{(\boldsymbol{Q}_{\mathrm{R}}-2\sigma_{\mathrm{w}}^2\boldsymbol{x}/D)^2}{4\sigma_{\mathrm{w}}^2/D}} \mathrm{e}^{-\frac{\boldsymbol{Q}_{\mathrm{I}}^2}{4\sigma_{\mathrm{w}}^2/D}+\mathrm{i}\boldsymbol{x}\cdot\boldsymbol{Q}_{\mathrm{I}}}\tilde{\varphi}(\boldsymbol{Q}). \tag{39}$$

It can be deduced from the above expression that the physical understanding of the GP field should be the combination of the space-time representation Gaussian waves and the momentum representation Gaussian wave packets, considering the variables' transformations $\boldsymbol{Q}_{\mathrm{R}} \to 2\sigma_{\mathrm{w}}^2\boldsymbol{y}/D$ and $\boldsymbol{Q}_{\mathrm{I}} \to \boldsymbol{q}$, the quantum field $\tilde{f}_{\mathrm{eff}}$ can be written as the direct product of the two Gaussian wave packets as

$$\begin{aligned} \tilde{f}_{\mathrm{eff}}(\boldsymbol{x}) = \sqrt{2}\sigma_{\mathrm{w}} &\int \mathrm{d}^D\boldsymbol{y}\, \psi_{\boldsymbol{p}=\boldsymbol{0}}(2^{-1}\sqrt{D}/\sigma_{\mathrm{w}}, \boldsymbol{x}, \boldsymbol{y})\hat{\kappa}(\boldsymbol{y}) \\ &\otimes \int \mathrm{d}^D\boldsymbol{q}\, \psi_{\boldsymbol{x}}(\sigma_{\mathrm{w}}/\sqrt{D}, \boldsymbol{q}, \boldsymbol{0})\hat{\eta}(\boldsymbol{q}), \end{aligned} \tag{40}$$

with $\langle\hat{\kappa}(\boldsymbol{x})\hat{\kappa}(\boldsymbol{y})\rangle = \delta^{(D)}(\boldsymbol{x} - \boldsymbol{y})$, $\langle\hat{\eta}(\boldsymbol{p})\hat{\eta}(\boldsymbol{q})\rangle = \delta^{(D)}(\boldsymbol{p} - \boldsymbol{q})$, and the wave function of a Gaussian wave packet being [44]

$$\psi_{\boldsymbol{p}}(\sigma, \boldsymbol{x}, \boldsymbol{x}_0) \equiv \frac{1}{(2\pi\sigma^2)^{D/4}}\mathrm{e}^{-\frac{(\boldsymbol{x}-\boldsymbol{x}_0)^2}{4\sigma^2}-\mathrm{i}\boldsymbol{p}\cdot(\boldsymbol{x}-\boldsymbol{x}_0)}, \tag{41}$$

$$\psi_{\boldsymbol{x}}(\sigma, \boldsymbol{q}, \boldsymbol{q}_0) \equiv \frac{1}{(2\pi\sigma^2)^{D/4}}\mathrm{e}^{-\frac{(\boldsymbol{q}-\boldsymbol{q}_0)^2}{4\sigma^2}+\mathrm{i}\boldsymbol{x}\cdot(\boldsymbol{q}-\boldsymbol{q}_0)}. \tag{42}$$

There exist a Fourier transformation relation that

$$\hat{\kappa}(\boldsymbol{y}) = \int \frac{\mathrm{d}^D\boldsymbol{q}}{(2\pi)^{D/2}}\mathrm{e}^{-\mathrm{i}\boldsymbol{y}\cdot\boldsymbol{q}}\hat{\eta}(\boldsymbol{q}), \tag{43}$$

$$\psi_{\boldsymbol{p}=\boldsymbol{0}}\left((2\sigma)^{-1}, \boldsymbol{x}, \boldsymbol{y}\right) = \int \frac{\mathrm{d}^D\boldsymbol{q}}{(2\pi)^{D/2}}\mathrm{e}^{-\mathrm{i}\boldsymbol{y}\cdot\boldsymbol{q}}\psi_{\boldsymbol{x}}(\sigma, \boldsymbol{q}, \boldsymbol{0}), \tag{44}$$

then both parts of the direct product are the same Gaussian wave packets, with the real part of $\boldsymbol{Q}$ being the space coordinate $2\sigma_{\mathrm{w}}^2\boldsymbol{y}/D$ while its imaginary part of $\boldsymbol{Q}$ being the momentum $\boldsymbol{q}$. The network's infinite-width quantum state is therefore a tensor product of identical Gaussian wave packets in phase space, consistent with the kernel derive above.

## 4. Fermionic Wick structure extension of NN-QFT

We now promote the network's hidden-to-output weights to *tensor* objects that carry an additional Clifford algebra index, thereby endowing the emergent field with fermionic Wick structured correlation functions.

According to the complex-valued scalar field theory, the free propagator will connect the mutually conjugate fields, the main problem is to construct the Grassman number, or the anticommute characters of fermionic fields. Construct the anticommuting matrices $\gamma_h$ via Pauli matrices. For $H = 3$, set $\gamma_h^{(3)} \equiv \sigma_h$ $(h = 1, 2, 3)$. For $H$ being odd numbers and $H \geq 5$, one can construct the gamma matrices by recursive definition as [45,46]

$$\gamma_h^{(H)} \equiv -\sigma_2 \otimes \gamma_h^{(H-2)},\ h = 1, 2, \ldots, H-2, \tag{45}$$

$$\gamma_{H-1}^{(H)} \equiv \sigma_1 \otimes \boldsymbol{I}_{\frac{d}{2}\times\frac{d}{2}}, \tag{46}$$

$$\gamma_H^{(H)} \equiv \sigma_3 \otimes \boldsymbol{I}_{\frac{d}{2}\times\frac{d}{2}}, \tag{47}$$

with the matrices $\gamma_h^{(H)}$ meet the Clifford algebra $\{\gamma_h^{(H)}, \gamma_{h'}^{(H)}\} = 2\delta_{hh'}I_{d\times d}$ and $(\gamma_h^{(H)})^\dagger = \gamma_h^{(H)}$, the dimensions of $\gamma_h^{(H)}$ should be $d = 2^{(H-1)/2}$. Motivated by the Section 3, only the last layer weight parameters control the spin character of the output field, with the other parameters being the eigenvalues of the quantum field, transform these weighting parameters to the tensor-formatted coefficients by multiplying the gamma matrices as

$$\varphi_h \to \varphi_h\gamma_h,\ \varphi_h^* \to \varphi_h^*\gamma_h^\dagger,\ h = 1, 2, \ldots, H. \tag{48}$$

Take the 4-pt correlation function as a example, there exist the trace calculation relation that

$$\begin{aligned} &\frac{1}{d}\left\langle\varphi_{h_1}^*\varphi_{h_2}^*\varphi_{h_1'}\varphi_{h_2'}\right\rangle\mathrm{tr}[\gamma_{h_1}\gamma_{h_2}\gamma_{h_1'}\gamma_{h_2'}] \\ &= \frac{4\sigma_{\mathrm{w}}^4}{H^2}(2\delta_{h_1h_2}\delta_{h_1h_1'}\delta_{h_1h_2'} - \delta_{h_1h_1'}\delta_{h_2h_2'} + \delta_{h_1h_2'}\delta_{h_2h_1'}), \end{aligned} \tag{49}$$

with the last two terms inside the bracket give the anticommute characters of $\varphi_h$.

The auxiliary-field construction used in the bosonic case cannot be transferred in a naive way to the present tensor-weighted setting: simply replacing the complex scalar field by a Clifford-matrix-valued field in the Gaussian auxiliary-field representation does not generate the desired tensor-weighted neural-network output. In particular, functional derivatives with respect to a scalar Grassmann source do not produce the extra Clifford indices carried by

$f(\boldsymbol{Q}, V|\boldsymbol{x})$, considering the inequation that

$$\frac{\delta}{\delta J(\boldsymbol{y})} e^{\int d^D \boldsymbol{x} J(\boldsymbol{x}) f_\gamma(\boldsymbol{Q}, V|\boldsymbol{x})} \Big|_{J=0} \neq f_\gamma(\boldsymbol{Q}, V|\boldsymbol{y}). \quad (50)$$

However, in the infinite width hidden layer condition, $H \to \infty$, higher-order vertex contributions are suppressed, and the remaining free-propagator contractions combined with Clifford-trace identities yield the alternating-sign pairing structure characteristic of fermionic Wick contractions for correlators with equal numbers of $f^\dagger$ and $f$. Note the statement in above concerns the large-width correlators of the matrix-valued output field rather than an algebraic convergence of Clifford generators to Grassmann variables.

## 5. Summary

This paper presents a theoretical framework that connects complex-valued neural networks (CVNNs) with tensor-valued output weights to fermionic correlation structures in the NN-QFT. For scalar output weights we first derive the generating functional of CVNNs, identify the exact Gaussian process that emerges in the infinite-width limit, and interpret the associated quantum state as a superposition of Gaussian wave packets in phase space. We then promote the last-layer weights to Clifford-algebra-valued tensors and show that, in the infinite-width limit and for correlators containing equal numbers of $f^\dagger$ and $f$, the resulting correlators obey fermionic Wick's theorem with respect to a scalar kernel. In this sense the tensor-weighted CVNN realizes a sign-structured surrogate for free Euclidean complex-fermion correlators at the level of Wick contractions and Feynman rules.

The reliance on independent Gaussian weight distributions and infinite-width approximations may oversimplify practical networks, where non-Gaussian fluctuations or finite-width effects could alter field-theoretic predictions. On the theoretical front, extending the framework to interacting fermionic systems (e.g., Yukawa or gauge theories) could deepen connections between deep learning and high-energy physics. Practically, the tensorization method provides a blueprint for encoding fermionic symmetries into neural architectures, potentially enhancing simulations of quantum many-body systems or lattice field theories. We stress that the present construction is primarily a theoretical extension of NN-QFT at the level of large-width correlator structure. The exponential growth of the Clifford representation dimension with H makes direct finite-width implementations challenging. Exploring reduced representations or alternative sign-generating architectures is an important direction for future work.

## Data availability

No data was used for the research described in the article.

## Declaration of competing interest

The authors declare the following financial interests/personal relationships which may be considered as potential competing interests: Kai Zhou reports financial support was provided by Shen- zhen Peacock fund. If there are other authors, they declare that they have no known competing financial interests or personal relationships that could have appeared to influence the work reported in this paper.

## Acknowledgements

The authors thank Profs. Pengfei Zhuang, Yi Yin, Xing Wang, Yao Ji, Jianhui Zhang and Bowen Xiao for helpful discussions. This work is supported by the CUHK-Shenzhen University development fund under grant No. UDF01003041 and UDF03003041, and Shenzhen Peacock fund under No. 2023TC0179, and NSFC fund under No. 92570117.

## Appendix A. Tensor weights 2n-points correlation function

In order to show the $H$ (last hidden layer's width) suppression to vertex terms in Lagrangian and prove the Fermionic Feynman rules for $2n$-points correlation functions, we denote $\phi_h \equiv \sqrt{H} * W_{\mathrm{h,out}}$, so that the distribution of $\phi_h$ is independent of $H$, and the corresponding $2n$-points correlation function with $n - f$ fields and $n$ conjugate transpose fields $f^\dagger$ can be represented by $\phi_h$ as

$$\begin{aligned}&\langle f^\dagger(\boldsymbol{Q}, V|\boldsymbol{x}_1) f^\dagger(\boldsymbol{Q}, V|\boldsymbol{x}_2) \cdots \\ &f^\dagger(\boldsymbol{Q}, V|\boldsymbol{x}_N) f(\boldsymbol{Q}, V|\boldsymbol{y}_1) f(\boldsymbol{Q}, V|\boldsymbol{y}_2) \cdots f(\boldsymbol{Q}, V|\boldsymbol{y}_N)\rangle \\ &= \frac{1}{H^N} \sum_{h_1 \cdots h_N = 1}^{H} \sum_{h'_1 \cdots h'_N = 1}^{H} \left\langle \prod_{i=1}^{N} \lambda^*_{h_i}(\boldsymbol{Q}, V|\boldsymbol{x}_i) \phi^*_{h_i} \prod_{j=1}^{N} \lambda_{h'_j}(\boldsymbol{Q}, V|\boldsymbol{y}_j) \phi_{h'_j} \right\rangle \\ &\frac{1}{d} \mathrm{Tr}\left[\gamma_{h_1}\gamma_{h_2} \cdots \gamma_{h_N} \gamma_{h'_1} \gamma_{h'_2} \cdots \gamma_{h'_N}\right],\end{aligned}$$

in the $H \to \infty$ limitation, after performing the fields' contraction and calculating the trace function, only in the condition that the number of the remaining $h_i$- and $h'_i$-summations large than $N$ will not be suppressed by the coefficient $1/H^N$, then only the free propagators condition (the contractions like $\langle \phi^*_h \phi_h \rangle$) will survive in this limitation, and all the values of $h'_j$ $(j = 1, 2, \cdots, N)$ should match the values of $h_i$ $(i = 1, 2, \cdots, N)$, with different orders, and all the values of $h_i$ $(i = 1, 2, \cdots, N)$ should be different, since the same value of $h_i$ will reduce the number of $h_i$- and $h'_i$-summations, therefore, for a given series of $h'_j$, one can assume $h'_j = h_{f(j)}$. Then in order to calculate the above equation to free propagator level, one can rearrange the order of $\gamma_{h_{f(1)}}\gamma_{h_{f(2)}} \cdots \gamma_{h_{f(N)}}$ first. Since all values of $h'_j (= h_{f(j)})$ are different, this procedure will give a plus or minus sign $\mathcal{P}_{h_{f(1)} h_{f(2)} \cdots h_{f(N)}}$, which depends on the odevity of the number of exchanging the order of list $\gamma_{h_{f(1)}}\gamma_{h_{f(2)}} \cdots \gamma_{h_{f(N)}}$ to $\gamma_{h_N} \cdots \gamma_{h_2}\gamma_{h_1}$, namely, $\mathcal{P}_{h_{f(1)} h_{f(2)} \cdots h_{f(N)}} = \mathcal{P}_{f(1) f(2) \cdots f(N)}$, with the right part of the equation is obtained by rearraging $f(1) f(2) \cdots f(N)$ to the order of $N, \cdots, 2, 1$.

$$\begin{aligned}&\langle f^\dagger(\boldsymbol{Q}, V|\boldsymbol{x}_1) f^\dagger(\boldsymbol{Q}, V|\boldsymbol{x}_2) \cdots \\ &f^\dagger(\boldsymbol{Q}, V|\boldsymbol{x}_N) f(\boldsymbol{Q}, V|\boldsymbol{y}_1) f(\boldsymbol{Q}, V|\boldsymbol{y}_2) \cdots f(\boldsymbol{Q}, V|\boldsymbol{y}_N)\rangle \\ &= \frac{1}{H^N} \sum_{h_1 \cdots h_N = 1}^{H} \sum_{\{f(j)\}} \mathcal{P}_{f(1) f(2) \cdots f(N)} \\ &\times \left\langle \prod_{i=1}^{N} \lambda^*_{h_i}(\boldsymbol{Q}, V|\boldsymbol{x}_i) \phi^*_{h_i} \prod_{j=1}^{N} \lambda_{h_{f(j)}}(\boldsymbol{Q}, V|\boldsymbol{y}_j) \phi_{h_{f(j)}} \right\rangle \\ &\times \frac{1}{d} \mathrm{Tr}\left[\gamma_{h_1}\gamma_{h_2} \cdots \gamma_{h_N} \gamma_{h_N} \cdots \gamma_{h_2}\gamma_{h_1}\right],\end{aligned}$$

with $\{f(j)\}$ represents different order conditions of $f(j)$. Considering that the last line of the above equation equals to 1 and assuming the inverse function of $i = f(j)$ being $j = f^{-1}(i)$, then the above expression can be transformed to be

$$\begin{aligned}&\langle f^\dagger(\boldsymbol{Q}, V|\boldsymbol{x}_1) f^\dagger(\boldsymbol{Q}, V|\boldsymbol{x}_2) \cdots \\ &f^\dagger(\boldsymbol{Q}, V|\boldsymbol{x}_N) f(\boldsymbol{Q}, V|\boldsymbol{y}_1) f(\boldsymbol{Q}, V|\boldsymbol{y}_2) \cdots f(\boldsymbol{Q}, V|\boldsymbol{y}_N)\rangle \\ &= \frac{1}{H^N} \sum_{h_1 \cdots h_N = 1}^{H} \sum_{\{f(j)\}} \mathcal{P}_{f(1) f(2) \cdots f(N)} \\ &\times \left\langle \prod_{i=1}^{N} \left[\lambda^*_{h_i}(\boldsymbol{Q}, V|\boldsymbol{x}_i) \phi^*_{h_i} \lambda_{h_i}(\boldsymbol{Q}, V|\boldsymbol{y}_{f^{-1}(i)}) \phi_{h_i}\right] \right\rangle,\end{aligned}$$

the two points connections $\phi^*_{h_i}\phi_{h_i}$ give the same result $2\sigma^2_{\mathrm{w}}$, and the expectations give the same value for arbitrary $h_i$, namely,

$$\begin{aligned}&\langle f^\dagger(\boldsymbol{Q}, V|\boldsymbol{x}_1) f^\dagger(\boldsymbol{Q}, V|\boldsymbol{x}_2) \cdots \\ &f^\dagger(\boldsymbol{Q}, V|\boldsymbol{x}_N) f(\boldsymbol{Q}, V|\boldsymbol{y}_1) f(\boldsymbol{Q}, V|\boldsymbol{y}_2) \cdots f(\boldsymbol{Q}, V|\boldsymbol{y}_N)\rangle \\ &= \frac{A^N_H}{H^N} \sum_{\{f(j)\}} \mathcal{P}_{f(1) f(2) \cdots f(N)} \left\langle \prod_{i=1}^{N} \left[\lambda^*_1(\boldsymbol{Q}, V|\boldsymbol{x}_i) \phi^*_1 \lambda_1(\boldsymbol{Q}, V|\boldsymbol{y}_{f^{-1}(i)}) \phi_1\right] \right\rangle,\end{aligned}$$

the coefficient $A^N_H = H!/(H - N)!$ comes from the number of permutation of choosing $N$ different numbers from numbers 1 to $H$, which comes from the degeneracy that for arbitrary $\{h_i\}$ series, the indexes' relation function $f(j)$ can be the same. In the $H \to \infty$ limit, $A_H^N/H^N \to 1$, finally, we get the result for $H \to \infty$,

$$\lim_{H\to\infty} \langle f^\dagger(\boldsymbol{Q}, V|\boldsymbol{x}_1) f^\dagger(\boldsymbol{Q}, V|\boldsymbol{x}_2) \cdots$$
$$f^\dagger(\boldsymbol{Q}, V|\boldsymbol{x}_N) f(\boldsymbol{Q}, V|\boldsymbol{y}_1) f(\boldsymbol{Q}, V|\boldsymbol{y}_2) \cdots f(\boldsymbol{Q}, V|\boldsymbol{y}_N)\rangle$$
$$= \sum_{\{f(j)\}} \mathcal{P}_{f(1)f(2)\cdots f(N)} \left\langle \prod_{j=1}^{N} \left[\lambda_1^*(\boldsymbol{Q}, V|\boldsymbol{x}_{f(j)})\phi_1^* \lambda_1(\boldsymbol{Q}, V|\boldsymbol{y}_j)\phi_1\right] \right\rangle,$$

which satisfies the Fermionic Feynman rules for representing $2n$-points diagrams with free propagators.

## References

[1] J. Halverson, A. Maiti, K. Stoner, Neural networks and quantum field theory, Mach. Learn. Sci. Tech. 2 (3) (2021) 035002. 2008.08601
[2] K.T. Grosvenor, R. Jefferson, The edge of chaos: quantum field theory and deep neural networks, SciPost Phys. 12 (3) (2022) 081. 2109.13247
[3] E. Dyer, G. Gur-Ari, Asymptotics of wide networks from Feynman diagrams (2019). 1909.11304
[4] S. Yaida, Non-Gaussian processes and neural networks at finite widths (2019). 1910.00019
[5] D. Bachtis, G. Aarts, B. Lucini, Quantum field-theoretic machine learning, Phys. Rev. D 103 (7) (2021) 074510. 2102.09449
[6] J.A. Zavatone-Veth, C. Pehlevan, Exact marginal prior distributions of finite Bayesian neural networks (2021). 2104.11734
[7] A. Maiti, K. Stoner, J. Halverson, Symmetry-via-duality: invariant neural network densities from parameter-space correlators (2021). 2106.00694
[8] D.A. Roberts, S. Yaida, B. Hanin, The Principles of Deep Learning Theory, Cambridge University Press, 2022. 2106.10165
[9] M. Demirtas, J. Halverson, A. Maiti, M.D. Schwartz, K. Stoner, Neural network field theories: non-Gaussianity, actions, and locality, Mach. Learn. Sci. Tech. 5 (1) (2024) 015002. 2307.03223
[10] J. Erdmenger, K.T. Grosvenor, R. Jefferson, Towards quantifying information flows: relative entropy in deep neural networks and the renormalization group, SciPost Phys. 12 (1) (2022) 041. 2107.06898
[11] H. Erbin, V. Lahoche, D.O. Samary, Non-perturbative renormalization for the neural network-QFT correspondence, Mach. Learn. Sci. Tech. 3 (1) (2022) 015027. 2108.01403
[12] D. García-Martín, M. Larocca, M. Cerezo, Quantum neural networks form Gaussian processes (2023). 2305.09957
[13] H. Erbin, V. Lahoche, D.O. Samary, Renormalization in the neural network-quantum field theory correspondence, 2022. 2212.11811
[14] A. Demichev, A. Kryukov, Interrelation of equivariant Gaussian processes and convolutional neural networks, J. Phys. Conf. Ser. 2438 (1) (2023) 012095. 2209.08371
[15] S.J. Du, G.K.L. Chan, Neuralized fermionic tensor networks for quantum many-body systems (2025). 2506.08329
[16] K. Choo, A. Mezzacapo, G. Carleo, Fermionic neural-network states for ab-initio electronic structure, Nature Commun. 11 (1) (2020) 2368.
[17] J. Kim, G. Pescia, B. Fore, J. Nys, G. Carleo, S. Gandolfi, M. Hjorth-Jensen, A. Lovato, Neural-network quantum states for ultra-cold Fermi gases, Commun. Phys. 7 (1) (2024) 148. 2305.08831
[18] A. Lovato, C. Adams, G. Carleo, N. Rocco, Hidden-nucleons neural-network quantum states for the nuclear many-body problem, Phys. Rev. Res. 4 (4) (2022) 043178. 2206.10021
[19] Z. Liu, B.K. Clark, Unifying view of fermionic neural network quantum states: from neural network backflow to hidden fermion determinant states, Phys. Rev. B 110 (11) (2024) 115124. 2311.09450
[20] G. Cassella, H. Sutterud, S. Azadi, N.D. Drummond, D. Pfau, J.S. Spencer, W.M.C. Foulkes, Discovering quantum phase transitions with fermionic neural networks, Phys. Rev. Lett. 130 (3) (2023) 036401. 2202.05183
[21] J.R. Moreno, G. Carleo, A. Georges, J. Stokes, Fermionic wave functions from neural-network constrained hidden states, Proc. Nat. Acad. Sci. 119 (32) (2022) e2122059119. 2111.10420
[22] Y. Nagai, M. Okumura, K. Kobayashi, M. Shiga, Self-learning hybrid Monte Carlo: a first-principles approach, Phys. Rev. B 102 (2020) 041124.
[23] G. Carleo, M. Troyer, Solving the quantum many-body problem with artificial neural networks, Science 355 (6325) (2017) 602–606.
[24] L. Wang, Discovering phase transitions with unsupervised learning, Phys. Rev. B 94 (2016) 195105.
[25] D. Wu, L. Wang, P. Zhang, Solving statistical mechanics using variational autoregressive networks, Phys. Rev. Lett. 122 (2019) 080602.
[26] H. Jiequn, L. Zhang, C. Roberto, E. Weinan, Deep potential: a general representation of a many-body potential energy surface, Commun. Comput. Phys. 23 (3) (2018) 629–639.
[27] K. Zhou, L. Wang, L.G. Pang, S. Shi, Exploring QCD matter in extreme conditions with machine learning, Prog. Part. Nucl. Phys. 135 (2024) 104084. 2303.15136
[28] L. Jiang, L. Wang, K. Zhou, Deep learning stochastic processes with QCD phase transition, Phys. Rev. D 103 (11) (2021) 116023. 2103.04090
[29] L. Wang, G. Aarts, K. Zhou, Diffusion models as stochastic quantization in lattice field theory, JHEP 05 (2024) 060. 2309.17082
[30] K. Zhou, G. Endrődi, L.G. Pang, H. Stöcker, Regressive and generative neural networks for scalar field theory, Phys. Rev. D 100 (1) (2019) 011501. 1810.12879
[31] K. Cranmer, J. Brehmer, G. Louppe, The frontier of simulation-based inference, Proc. Natl. Acad. Sci. 117 (48) (2020) 30055–30062.
[32] G. Aarts, D.E. Habibi, L. Wang, K. Zhou, On learning higher-order cumulants in diffusion models, Mach. Learn. Sci. Tech. 6 (2) (2025) 025004. 2410.21212
[33] S. Chen, O. Savchuk, S. Zheng, B. Chen, H. Stoecker, L. Wang, K. Zhou, Fourier-flow model generating Feynman paths, Phys. Rev. D 107 (5) (2023) 056001. 2211.03470
[34] J. Chen, W.-K. Wong, B. Hamdaoui, A. Elmaghbub, K. Sivanesan, R. Dorrance, L.L. Yang, An analysis of complex-valued cnns for rf data-driven wireless device classification, in: ICC 2022-IEEE International Conference on Communications, IEEE, 2022, pp. 4318–4323.
[35] C. Lee, H. Hasegawa, S. Gao, Complex-valued neural networks: a comprehensive survey, IEEE/CAA J. Autom. Sin. 9 (8) (2022) 1406–1426.
[36] J.A. Barrachina, C. Ren, G. Vieillard, C. Morisseau, J.P. Ovarlez, Theory and implementation of complex-valued neural networks, arXiv:2302.08286 (2023).
[37] J. Bassey, L. Qian, X. Li, A survey of complex-valued neural networks, 2021. arXiv:2101.12249
[38] C. Trabelsi, O. Bilaniuk, Y. Zhang, D. Serdyuk, S. Subramanian, J.F. Santos, S. Mehri, N. Rostamzadeh, Y. Bengio, C.J. Pal, Deep complex networks, 2018. arXiv:1705.09792
[39] J.S. Dramsch, M. Lüthje, A.N. Christensen, Complex-valued neural networks for machine learning on non-stationary physical data, Comput. Geosci. 146 (2021) 104643. .
[40] J.H. Wu, S.Q. Zhang, Y. Jiang, Z.H. Zhou, Complex-valued neurons can learn more but slower than real-valued neurons via gradient descent, in: A. Oh, T. Naumann, A. Globerson, K. Saenko, M. Hardt, S. Levine (Eds.), Advances in Neural Information Processing Systems, 36, Curran Associates, Inc., 2023, pp. 23714–23747.
[41] F. Voigtlaender, The universal approximation theorem for complex-valued neural networks, Appl. Comput. Harmon. Anal. 64 (2023) 33–61.
[42] M.E. Peskin, D.V. Schroeder, An Introduction to Quantum Field Theory, Addison-Wesley, Reading, USA, 1995.
[43] R. Shankar, Renormalization group approach to interacting fermions, Rev. Mod. Phys. 66 (1994) 129–192. cond-mat/9307009
[44] S. Olivares, Quantum optics in the phase space: a tutorial on Gaussian states, The European Physical Journal Special Topics 203 (1) (2012) 3–24.
[45] R. Davies, Construction of spinors in various dimensions, 2011. Unpublished lecture notes, http://rhysdavies.info/physics_page/resources/notes/spinors.pdf.
[46] J.H. Park, Lecture note on Clifford algebra, J. Korean Phys. Soc. 81 (2022) 1. 2205.09509